# Decoupled trends for electrical and thermal conductivity in phase-confined CNT co-continuous blends

by *S. Colonna[a], Z. Han[b], A. Fina\*[a]*

[a] Politecnico di Torino, Dipartimento di Scienza Applicata e Tecnologia, Viale Teresa Michel 5, 15121, Alessandria, Italy

[b] College of Materials Science and Engineering, Harbin University of Science and Technology, Harbin, China

Email: alberto.fina@polito.it

**Abstract**

In the present work, the morphology and the electrical and thermal conduction properties of co-continuous poly(vinylidene fluoride) (PVDF), maleated polypropylene (PPgMA) and multiwall carbon nanotubes (CNT) nanostructured blends are investigated. CNT preferentially locates in the PPgMA phase and clearly causes a refinement in the co-continuous structure. Electrical conductivity experiments show that nanocomposites are well above the percolation threshold and evidence for one order of magnitude enhancement in conductivity for the co-continuous nanocomposites compared to the monophasic nanocomposites with the same CNT volume fraction. On the other hand, thermal diffusivity enhancement for the co-continuous blends is found lower than that for the monophasic nanocomposites at the same CNT volume fraction. An explanation is proposed in terms of large interfacial area, causing phonon scattering at the interface between immiscible PVDF and PPgMA domains. Results described in this paper open the way to the preparation of high electrical and low thermal conductivity materials with possible application as thermoelectrics.

## 1. Introduction

Thermal management of polymeric materials is of high interest in energy related domains for thermal energy dissipation, including electronics, low temperature heat exchange and recovery, thermal storage, flexible heat spreader, encapsulation layers, etc. [1, 2, 3]. On the other hand, electrical conductivity of polymeric materials is also of utmost interest for application in electromagnetic shielding, conductive coatings and thermoelectrics [4, 5, 6]. As common polymers are both thermally and electrically insulators, conductive particles, including graphite, metal or ceramic particles, as well as nanoparticles such as carbon nanotubes, graphene and boron nitride, are typically exploited to improve the thermal and electrical conductivity of insulating matrix. While in most cases the enhancement of electrical conductivity is obtained along with a thermal conductivity enhancement, decoupling of thermal and electrical conductivity is needed in some applications. As an example, some heat management applications in electronics need high thermal conductivity coupled with electrical insulation, which is typically obtained with ceramic nanoparticles, such as boron nitride or aluminum nitride [7, 8, 9]. On the other hand, the enhancement of electrical conductivity typically brings a certain increase in the thermal conductivity of the material. This represents a problem in thermoelectric materials, in which efficiency is dependent on the figure of merit, $ZT = S^2\sigma T/\kappa$, where $S$ is the Seebeck coefficient, σ is the electrical conductivity, $\kappa$ is the thermal conductivity and T is the temperature [4, 10]. To enhance $ZT$, the material should therefore ideally have high electrical conductivity and low thermal conductivity.

Among conductive nanoparticles for the enhancement of charge and heat transfer in polymers, carbon nanotubes (CNT) attracted lot of interests since their discovery in 1991 [11], owing to their outstanding mechanical [12, 13], thermal [2, 14, 15] and electrical properties [14], for the improvement of the low thermal (0.2 - 0.5 W m$^{-1}$ K$^{-1}$ [2]) and negligible electrical conductivities (< 10$^{-10}$ S cm$^{-1}$ [14]) of polymer matrices. Indeed, dramatic enhancement in the electrical conductivity is generally obtained at low CNT content [15], with the percolation threshold, *i.e.* the transition from non-conducting to conducting material, occurring at CNT content as low as 0.002 wt.%, depending on the nanotube aspect ratio, degree of alignment, state of aggregation, as well as mixing condition and the type of polymer matrix [16, 17, 18, 19, 20, 21]. On the other hand, the improvement of thermal conductivity with the inclusion of CNT or other nanoparticles is significantly less sharp and strongly dependent on many factors, including nanoparticles type, quality, dispersion, alignment, contact between nanoparticles and interfacial interaction with the polymer matrix [2, 15, 17, 22, 23, 24, 25]. A possible route to enhance the effectiveness of a percolative network is the confinement of conductive particles in well-defined continuous regions [2, 26, 27, 28]. This might maximize the local density and therefore the contacts between particles into a dense percolation network while keeping a limited overall particle concentration. A possible solution to obtain particles confinement is the melt blending of two immiscible polymers in appropriate volume fractions aiming at the formation of two continuous phases interpenetrating each other, usually referred to as co-continuous structure [29, 30, 31, 32]. The selective localization of conductive fillers in one of the continuous phases depends on the interaction between the filler and the polymer matrix (thermodynamic factors), the melt-viscosity ratio of the constituent components and the processing parameters employed during compounding (kinetic factors) [33]. The inclusion of CNT into immiscible polymer blends showing co-continuous morphology was reported by several authors, typically showing preferential segregation of CNT in one of the phases [33, 34, 35, 36, 37, 38, 39]. Pötschke *et al*. [34] added CNT to a polyethylene (PE)/PC and measured the electrical percolation threshold at 0.41 vol.% CNT content. Studying acrylonitrile-butadiene-styrene (ABS)/PP/CNT nanocomposites, Khare *et al*. [35] observed a refinement of the co-continuous morphology induced by CNT and measured the electrical percolation threshold at 0.45 wt.% CNT content with PP/ABS ratio of 45/55. Furthermore, in recent works, researchers were able to control the segregation of CNT at the interface of immiscible polymer blends, by tuning the thermodynamics of the system [38] or the kinetics of the process [40], leading to electrical percolation threshold at 0.017 wt.% [38] and 0.025 wt.% [40], respectively. While electrical conductivity of double percolated blends was largely studied, the thermal conductivity of co-continuous polymer blends containing CNT has not been reported in literature, to the best of the authors' knowledge. However, we previously reported that the addition of graphite positively affected the thermal conductivity of polyvinylidenfluoride (PVDF)/maleated polypropylene (PPgMA) co-continuous polymer composites, with 60% enhancement in the thermal diffusivity of PVDF/PPgMA/Graphite respect to PPgMA/Graphite composites at the same graphite loading [26]. In this study, the morphology evolution of PVDF/PPgMA/CNT nanocomposites with nanotubes content is addressed together with the evaluation of the electrical and thermal conduction properties of the nanocomposites.

## 2. Experimental

### 2.1. Materials

PVDF (Solef® S1010) with MFI = 2 g/10min (230 °C, 2.16 kg), was purchased from Solvay (Belgium). PPgMA (Polibond® 3200) with maleic anhydride content = 1.0 wt. % and MFI = 115 g/10 min (190 °C, 2.16 kg), was purchased from Crompton (USA). CNT (NC-7000) with a carbon purity of 90% and an average diameter of 9.5 nm were purchased from Nanocyl (Belgium). All the materials were used as received.

### 2.2. Sample preparation

Nanocomposites were prepared by using a twin-screw micro-compounder (Xplore by DSM, Netherlands) with 15 cm$^3$ mixing chamber, a recirculating channel and two co-rotating conical screw. PVDF and PPgMA were first fed into the compounder and blended at 210°C for 3 minutes, then MWCNT were added mixing for further 5 minutes. The screw speed was set at 100 rpm and nitrogen flow was used to prevent thermal oxidation during melt-mixing. Compounds were prepared varying the CNT fraction at 3.4 vol.%; 5.1 vol.%; 6.8 vol.% and 10.4 vol.% .

The as prepared nanocomposites were later molded at 220 °C and 1 min and 100 bar, by means of a laboratory press (Gibitre instruments, Italy).

### 2.3. Morphological characterization

The morphology of the nanocomposites was observed by scanning electron microscopy (SEM), with a LEO 1450 VP instrument (France) coupled with a back scattered electron detector and an EDS elemental analysis INCA Energy 7353 (Oxford Instruments, UK) probe. Samples for SEM observation where cut at cryo temperature (-30°C) with a microtome to obtain planar and smooth surfaces. The as prepared samples were then gold coated to ensure electrical conduction from sample surface. Transmission electron microscopy was carried out on a Jeol 2010 Field Emission TEM at 100 kV. Observations were performed on 60 nm thick slices ultracryocut parallel to the extrusion direction at -60 °C.

### 2.4. Thermal diffusivity analysis

Thermal diffusivity was measured by LFA 427 (Laser-Flash-Apparatus, Netzsch, Germany) on samples with 12.5 mm and 1 mm in diameter and height, respectively. Measurements were repeated three times on three different specimens for each nanocomposite to evaluate the standard deviation.

### 2.5. Electrical resistivity analysis

Volume resistivity measurements were carried out on thermal diffusivity specimens by a home-made apparatus composed by a regulated power supply (GPS-3303 by GW Instek, Taiwan), two multimeters equipped with a digital filter to reduce the measurement noise (8845A by Fluke, USA) and two homemade electrodes. One multimeter was used for the regulation of the electric current while the other for voltage measurements. The electrodes were a cylinder (18.5 mm diameter, 55 mm height) and a square plate (100 mm side, 3mm thickness), both made of brass. Every electrode has a hole for the connection and a wire furnished with a 4 mm banana plug. The measurement system works with the multimeter method. The power supply was time to time regulated in current or in voltage to have accurate measurement by both the multimeters limiting, however, the power dissipated on the specimen. Before measurement, specimens were silver coated on both sides to improve the electrical conduction between them and the electrodes. Test were carried out on disks with 12.5 mm diameter and 1 mm thickness.

## 3. Results and discussion

### 3.1. Composites morphology

The formation of co-continuous structures in immiscible polymer blends requires the exploitation of defined volume ratio between the polymers constituting the phases [29]. In our previous work, we observed co-continuity in PVDF/PPgMA blends when the amount of PPgMA was between 30 and 40% by weight, despite a coarse phase separation sizing in the range of tens to a few hundreds of microns was observed [26]. Thus, in the present work, nanocomposites were prepared fixing the PVDF/PPgMA weight ratio at 7/3, which corresponds to about 54/46 ratio in volume, and varying the amount of CNT. The morphology of the nanocomposites was investigated by means of SEM. Results are showed in Figure 1, where PVDF appears as the light gray phase (in back-scattered signal, due to the presence of fluorine which is heavier than hydrogen) whereas PPgMA corresponds to the darker phase.

While all the formulations display the typical morphology of co-continuous blends, it is worth observing that increasing the amount of CNT leads to a refinement of the phase separation. Indeed, the mean thickness of the two phases decreases from some tens of micrometers for the nanocomposites containing 3.4 vol.% of CNT (Figure 1a) down to few micrometers for the nanocomposites containing 5.1 vol.% CNT (Figure 1b). These values are much lower than the hundreds of micrometers observed for pure PVDF/PPgMA 7/3 blends reported in our previous work [26]. This refinement of the co-continuous structure was observed also by other authors in the presence of nanoparticles [41, 42], and was related to the rise in the viscosity of the phase containing CNT, the reduction of the interfacial energy or the inhibition of coalescence by the presence of a solid barrier around the minor phase drops [43]. Furthermore, the combination of the yield stress of the nanocomposite polymer phase, the bending resistance of the nanoparticle and the interfacial tension between the two polymer phases was also proposed by Nuzzo et al. as a single numerical criterion to rationalize the formation of a cocontinuos nanocomposite blend [44]. In our case, TEM analysis (Figure 1c) reveals that CNTs are mostly located in the PPgMA phase, in accordance with the results observed with natural graphite in our previous work [26]. Confinement of CNT in PPgMA corresponds to a significantly higher local concentration of CNT. In particular, based on the 54/46 volume composition of the blend the local concentration of CNT in the PPgMA host phase (Table 1) in significantly higher compared to the nominal overall CNT concentration.

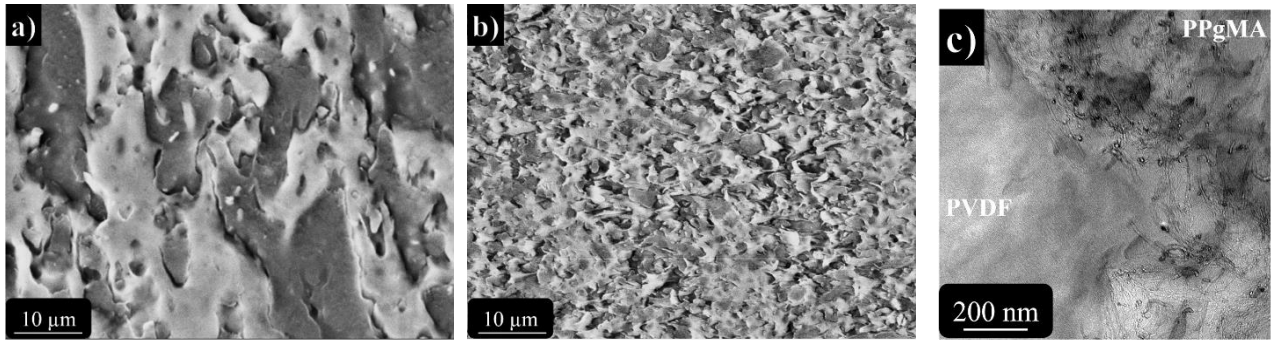

**Figure 1.** Morphology of PVDF/PPgMA/CNT nanocomposites: a) SEM micrograph for PVDF/PPgMA with 3.4 vol.% CNT; (b) SEM and (c) TEM micrographs for PVDF/PPgMA with 6.8 vol.% CNT

*3.2. Conductive properties of PVDF/PPgMA/CNT nanocomposites*

The segregation of electrically conductive nanoparticles into one phase of co-continuous polymer blends, is expected to result in the formation of the so-called double percolating network [45, 46]. The electrical conductivity results are reported in Figure 2a and summarized in Table 1. The results indicate that all the nanocomposites show the electrical conductivity in the range of $10^0$ to $10^1$ S/m, thus the CNT content is well above the percolation threshold, as expected for the relatively high contents of CNT. Monophasic PPgMA/CNT nanocomposites exhibit *approx.* one order of magnitude lower electrical conductivity respect to the co-continuous nanocomposites at given CNT volume fraction. For instance, at 10.4 vol.% CNT content, ~ 2.3 S m$^{-1}$ and ~ 16.5 S m$^{-1}$ are obtained for PPgMA/CNT and PVDF/PPgMA/CNT, respectively, thus confirming the positive effect of nanoparticle confinement in a continuous phase. The volume electrical conductivity enhancement is in principle expected to depend on both the percolation density and the volume fraction over which the percolation network is extended. However, in the present case, the local concentration of CNT in the host phase appears to primarily determine the enhancement of electrical conductivity for co-continuous nanocomposites compared to monophasic counterparts. For instance, by the comparison of PVDF/PPgMA/5.1%CNT and PPgMA + 10.4% CNT, the local concentration of CNT in the host phase (11.1% *vs.* 10.4%, respectively) is directly related to the comparable electrical conductivities obtained, respectively 4.3 ± 0.8 S m$^{-1}$ and 2.3 ± 0.9 S m$^{-1}$. The confinement of thermally conductive particles into one phase of the continuous polymer blends is also expected to result in enhanced thermal conductivity owing to the maximised particle-particle contact [26]. However, the results (Figure 2b and Table 1) reveal the thermal diffusivity for PVDF/PPgMA/CNT to be lower than monophasic PPgMA/CNT with the same CNT content. The highest value in Table 1, α ≈ 0.286 mm$^2$s$^{-1}$, is indeed measured for PPgMA based nanocomposite containing 10.4 vol.% CNT, whereas PVDF/PPgMA/CNT nanocomposite at the same volume fraction of CNT shows a thermal diffusivity of ≈ 0.247 mm$^2$s$^{-1}$. This result appears counter-intuitive and suggests the confinement of CNT in co-continuous blends may not be effective in improving thermal conductivity, in contrary with electrical conductivity. To understand these results, it is worth observing that also the thermal diffusivity of the neat PVDF/PPgMA blend (0.106 mm$^2$ s$^{-1}$) is lower than those of both the neat polymer matrices (0.118 and 0.121 mm$^2$ s$^{-1}$ for PPgMA and PVDF, respectively). This is explained by the presence of interfaces between the two polymer phases in the co-continuous morphology acting as additional thermal resistances in the heat transfer within the material. The interface between immiscible polymers is clearly associated to poor chemical interactions and possibly to microcracks at the interface, both phenomena resulting in a lower efficiency in the transfer of phonons than in bulk polymers. It was previously reported both experimentally and by theoretical models that, with the presence of particles much smaller than the Kapitza radius and/or totally insulating interfaces, the thermal conductivity of composites can be lower than that of the polymer matrix despite the higher intrinsic thermal conductivity of the particles [47]. In the case of co-continuous blends containing CNT, the confinement of CNT in the PPgMA phase clearly has increased its electrical conductivity, but also caused a refinement in the co-continuous structure, thus leading to an increase in the specific interfacial area, which opposes to the enhancement in thermal conductivity related to the presence of nanotubes. Overall, the advantage from CNT confinement is negatively compensated by the effect of larger interfacial area, causing in the lack of improved performance for the whole blend, compared to monophasic nanocomposites with the same CNT volume fraction.

The results obtained with CNT in PVDF/PPgMA blend clearly differ from the previously reported equivalent co-continuous blends containing graphite [26], where both electrical and thermal conductivity enhancements were obtained via confinement of graphite in the PPgMA phase compared with the monophasic composite. The difference between the effects CNT and graphite in the thermal conductivity may be explained by two reasons. First, a difference is obtained in the orientation of fillers: while the rigid graphite microparticles are easily oriented within the micrometer-size PPgMA phase, highly flexible CNT are not significantly oriented, as observable by TEM. Second, the effect of particles segregation within PPgMA phase appears to be more effective for platelet-like particles compared to tubular nanoparticles, in terms of higher extension of contact area, thus resulting in a more significant decrease in the contact resistance within the micro-graphite percolating network.

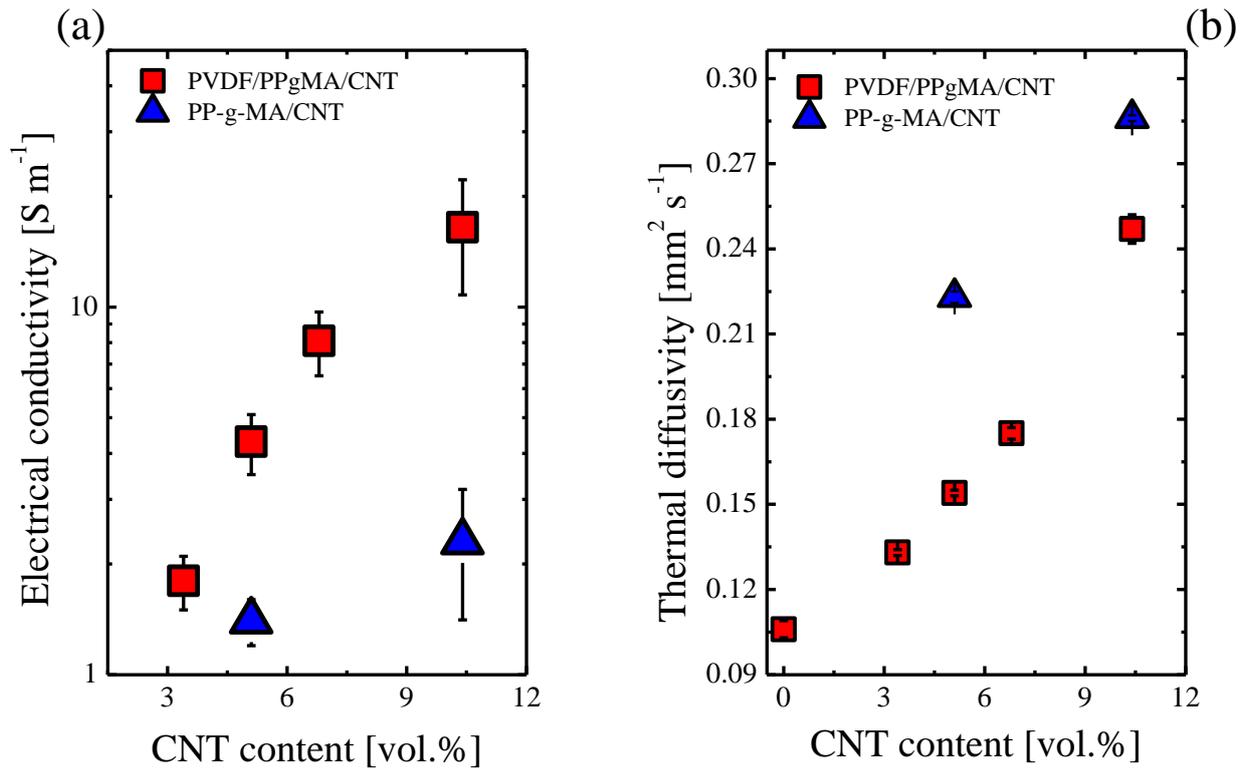

**Figure 2.** (a) Volumetric conductivity and (b) thermal diffusivity for PVDF/PPgMA/CNT and PPgMA/CNT nanocomposites.

**Table 1.** Thermal diffusivity and electrical conductivity values for PVDF/PPgMA/CNT and PPgMA/CNT nanocomposites.

| Samples | CNT content (vol.%) | CNT in the host phase (vol.%) | Thermal diffusivity ($mm^2 s^{-1}$) | Electrical conductivity ($S\ m^{-1}$) |
|---|---|---|---|---|
| PVDF/PPgMA | - | - | 0.106 ± 0.003 | $10^{-16}$ |
| PVDF/PPgMA/3.4%CNT | 3.4 | 7.4 | 0.133 ± 0.001 | 1.8 ± 0.3 |
| PVDF/PPgMA/5.1%CNT | 5.1 | 11.1 | 0.154 ± 0.001 | 4.3 ± 0.8 |
| PVDF/PPgMA/6.8%CNT | 6.8 | 14.8 | 0.175 ± 0.002 | 8.1 ± 1.6 |
| PVDF/PPgMA/10.4%CNT | 10.4 | 22.6 | 0.247 ± 0.005 | 16.5 ± 5.7 |
| PPgMA + 5.1%CNT | 5.1 | 5.1 | 0.223 ± 0.002 | 1.4 ± 0.2 |
| PPgMA + 10.4%CNT | 10.4 | 10.4 | 0.286 ± 0.001 | 2.3 ± 0.9 |

## 4. Conclusion

In the present paper, PVDF/PPgMA/CNT nanocomposites, with PVDF/PPgMA ratio of 70/30 by weight, were prepared through melt mixing in a micro compounder. Morphological studies on the as prepared nanocomposites demonstrate that the all the materials exhibit co-continuous morphology, with CNT mainly locates in the PPgMA phase and progressive refinement of the co-continuous structure with the increase of CNT content.

The electrical conductivity experiments reveal that all the nanocomposites are well above the percolation threshold. Furthermore, the confinement of CNT in the PPgMA phase, due to the formation of the so-called double percolating network, positively affect the electrical conductivity, being the electrical conductivity of the co-continuous nanocomposite higher (about one order of magnitude) than that of single matrix nanocomposite containing the same CNT volume fraction. On the other hand, thermal diffusivity enhancement for the co-continuous blends was found lower than for the monophasic nanocomposites at the same CNT volume fraction. This observation was interpreted in terms of concurrent effects between the positive effect of enhancement by CNT confinement in continuous phase and the negative effect of large interfacial area by CNT refinement on co-continuous structure. The phonon scattering at the interface between immiscible PVDF and PPgMA domains opposes to the enhancement in thermal conductivity related to the presence of nanotubes.

The results described in this paper provide more insight in the understanding of electrical and thermal conductivity in polymer blends and may open possible application in thermoelectric materials, where the coupling of high electrical conductivity and low thermal conductivity is required.

**Acknowledgements**

The research leading to this results was initially funder from the European Community's Seventh Framework Program under grant agreement no. 227407-Thermonano. More recently, the work was finalized with partial funding from the European Research Council (ERC) under the European Union's Horizon 2020 research and innovation programme under grant agreement 639495—INTHERM—ERC-2014-STG.

A.F. conceived this research work and the experiments within, interpreted the experimental results and led the research activities. S.C. carried out materials preparation and most of the characterization. Z.H. contributed to interpretation of results and discussion. The manuscript was mainly written by S.C and A.F.